\documentclass[pre,showpacs]{revtex4}

\usepackage{graphicx}

\begin{document}

\title{Joint Probability Distributions for a Class of
Non-Markovian Processes}

\author{A. Baule and R. Friedrich}

\affiliation{Institute of Theoretical Physics, Westf\"alische Wilhelms-Universit\"at M\"unster,
Wilhelm-Klemm-Str. 9, 48149 M\"unster, Germany}

\begin{abstract}
We consider joint probability distributions for the class of
coupled Langevin equations introduced by Fogedby 
[H.C. Fogedby, Phys. Rev. E 50, 1657 (1994)].
We generalize well-known results for the single
time probability distributions to the case of N-time joint probability
distributions. It is shown that these
probability distribution functions can be obtained by an
integral transform from distributions of a Markovian process.
The integral kernel obeys a partial differential equation with fractional time
derivatives reflecting the non-Markovian character of the process.

\end{abstract}
\pacs{
02.50.-r,
05.40.-a,
47.27.-i,
05.30.Pr
}

\maketitle

\section{Introduction}

In recent years, the connections between 
"continuous time random walk" (CTRW), 
which originated in the work of Montroll and Weiss
\cite{Weiss} 
generalizing the idea of Brownian random walks, and fractional
Fokker-Planck equations 
have been established. For a review we refer the reader to \cite{Metzler_rev}.
The solutions of these equations
exhibit both
super- and subdiffusive behaviour and are thus appropriate 
models for a large variety of transport processes in complex systems 
\cite{Bouchaud}. Recently, a connection between the velocity
increment statistics of a Lagrangian tracer particle in fully
developed turbulent flows and a type of CTRW
has been introduced \cite{Friedrich}. Here, a closure
assumption on a hierarchy of joint velocity-position pdf's 
derived from a statistical formulation of the Navier-Stokes equation
leads to a generalization of 
Obukhov's random walk model \cite{Obukhov} in terms of a continous
time random walk. It allows for a successful parametrization of the single
time probability distributions of velocity increments. However, there
are different suggestions for the stochastic process of Lagrangian
particles in turbulence, which are able to provide reasonable
approximations for the single time velocity increment
statistics. This example evidences that one has to introduce further
quantities in order to distinguish between different stochastic 
models.

For non-Markovian processes, the natural extension is the
consideration
of N-times joint probability distributions.
It seems that for the class of CTRWs
only single time probability distributions have been
investigated so far. In that case fractional diffusion equations of
the form 
\begin{equation}\label{frag}
\frac{\partial }{\partial t}f(x,t) =\:_0D_t^{1-\alpha}\: L\: f(x,t)
\end{equation}
can be derived.
Here x denotes the random variable, $L$ is a Fokker-Planck operator
(for diffusion processes 
$L=\frac{\partial^2 }{\partial x^2}$) and $_0D_t^{1-\alpha}$ is the
Riemann-Liouville fractional differential operator (c.f. appendix A).
The properties
of this equation with regard to physical applications have been extensively
discussed in the recent reviews \cite{Metzler_rev}, \cite{Metzler_rev2}.
In \cite{Fogedby} Fogedby introduced a class of
coupled Langevin equations, where he also considered a case which
leads to an operator $L$ including fractional derivatives with respect
to the variable x, 
$L=\frac{\partial^\beta }{\partial x^\beta}$.  A similar case
has been studied by Meerschaert et al. \cite{MeerPRE}, 
who made an extension to
several dimensions introducing a multidimensional generalization of 
fractional diffusion, so-called operator L\'evy motion. This allows
for a description of anomalous diffusion with direction dependent Hurst
indices $H_i$ defined by the relation $<(x_i(t)-x_i(t=0))^2> \approx t^{2
H_i}$. In \cite{MeerJAP} limit theorems of a class of continuous time
random walks with infinite mean waiting times have been
investigated. It is shown that the limit process obeys a fractional
Cauchy problem. The emphasis again is put on single time distributions.

The purpose of the present
paper is to investigate multiple time probability distribution
functions for the class of coupled Langevin equations introduced by
Fogedby \cite{Fogedby}, which have been considered to be a
representation of a continuous time random walk.\\
\\
The paper is outlined as follows. 
In the next section we present
the coupled Langevin equations considered by Fogedby
\cite{Fogedby} consisting of a usual Langevin process  
$X(s)$ in a coordinate $s$
and a L\'evy
process representing a stochastic relation $t(s)$. One is interested in
the process $X(t)=X(s^{-1}(t))$. Fogedby \cite{Fogedby} 
investigated the case where the processes $X(s)$ and $t(s)$ are
statistically independent and showed how fractional diffusion
equations of the form (\ref{frag}) arise. Interesting results for the
case where the processes are statistically dependent have been 
considered by Becker-Kern et al. \cite{MeerAP}
leading to generalizations of the fractional diffusion
equations (\ref{frag}). However, both
publications are devoted to single time probability distributions.

In section II we present a central formula, which relates the
N-times probability distributions of $X(t)$ to the pdf's of
$X(s)$ via an integral transform, which is determined by the
process $t(s)$. In section III properties of the involved
L\'evy-stable process $t(s)$ 
are considered leading to expressions for the pdf of
the inverse process $s(t)$.
In section V we specify the moments for the case of a  simple 
diffusion process.

\section{A Class of Non-Markovian Processes}

Starting point of our discussion is the set 
of coupled Langevin equations \cite{Fogedby} for the
motion of a Brownian particle in an external force field 
$F$ in d=1 dimensions
(an extension to higher dimensions $d>1$ is straightforward):

\begin{eqnarray}
\frac{dX(s)}{ds}&=& F(X) + \eta(s), \label{Langevin1}\\
\frac{dt(s)}{ds}&=& \tau(s)\qquad.	\label{Langevin2}	
\end{eqnarray}

In this framework the random walk is parametrized in terms of the 
continuous path
variable $s$, which may be considered eg. as arc length along the 
trajectory. 
$X(s)$ and $t(s)$ denote the position and time in physical space. 
The random variables $\eta(s)$ and $\tau(s)$ are responsible
for the stochastic character
of the process. We are only considering the case of uncoupled jump lengths
and waiting times such that $\eta$ and $\tau$ are statistically
independent (coupled CTRWs have been considered in \cite{MeerAP}). The arc
lenght is related to physical time $t$ by the inverse function
$s=t^{-1}(t)=s(t)$. Thus, we have to assume $\tau(s) > 0$.
We are interested in the process $X(s(t))$, i.e. the behaviour of the
variable X as a function of physical time t. \\
For the characterization of the process 
we introduce the two-times probability density functions (pdf)
for the processes (\ref{Langevin1}),
(\ref{Langevin2}):
\begin{eqnarray}
f_1(x_2,s_2;x_1,s_1)&=&<\delta(x_2-X(s_2))\delta(x_1-X(s_1))> \qquad 
,\label{pdf1}\\
p(t_2,s_2;t_1,s_1)&=&<\delta(t_2-t(s_2))\delta(t_1-t(s_1))>
\qquad , \label{pdf2}\\
f(x_2,t_2;x_1,t_1)&=&<\delta(x_2-X(s(t_2)))\delta(x_1-X(s(t_1)))> \qquad .
\label{pdf3}
\end{eqnarray}
Here the brackets $<..>$ denote a suitable average over
stochastic realizations. For the sake of simplicity
we restrict ourselves to n = 2. The generalization to multiple times
is obvious. Both probability 
functions are determined by the statistics of the independent
random variables $\eta$ and $\tau$. 

\subsection{The process $X(s)$}

We consider the case where $\eta(s)$ is the standard Langevin force,
i.e. $\eta$ is a Wiener process. 
In turn
(\ref{Langevin1}) becomes Markovian and
$f_1(x_2,s_2;x_1,s_1)$ can be determined
by solving the corresponding Fokker-Planck equation (FPE) for the
conditional probability distribution $P(x_2,s_2\mid x_1,s_1)$:
\begin{eqnarray}
\label{FPE}
\frac{\partial}{\partial s}P(x_2,s_2\mid x_1,s_1)
&=&\left(-\frac{\partial}{\partial x}F(x)+
\frac{\partial^2}{\partial x^2}\right)P(x_2,s_2\mid x_1,s_1)\nonumber\\
&=& L_{FP}(x) P(x_2,s_2\mid x_1,s_1) \qquad.
\end{eqnarray}
The diffusion constant is set to $1$ in the
following. Due to the Markovian property of the process 
$X(s)$ the joint pdf is obtained 
by multiplication with the single time pdf according to
\begin{eqnarray}
f_1(x_2,s_2;x_1,s_1)=P(x_2,s_2\mid x_1,s_1)f(x_1,s_1) \qquad.
\end{eqnarray}
For a general treatment of the FPE we refer the reader to 
the monographs of Risken \cite{Risken} and Gardiner \cite{Gardiner}.

\subsection{The process $t(s)$}

The stochastic process $t(s)$ is determined by the properties of
$\tau(s)$. The corresponding pdf's are denoted by $p(t,s)$,
$p(t_2,s_2;t_1,s_1)$.
Furthermore, we shall consider
$\tau(s)$ to be a (one-sided) L\'evy-stable process of order 
$\alpha$ \cite{Fogedby}, \cite{Schertzer} 
with $0<\alpha<1$.
As a result, the process $t(s)$ is Markovian. 
L\'evy-stable processes of this kind
induce the property of a diverging characteristic
waiting time $<t(s)>$ .
Consequently the stochastic process
in physical time $t$, given by the coupling of the Langevin
equations (\ref{Langevin1}) and (\ref{Langevin2})
reveals subdiffusive behaviour. The specific
form of $p(t_2,s_2;t_1,s_1)$ will be given below.\\
For a deeper discussion we refer to the 
review articles \cite{Metzler_rev}, \cite{Metzler_rev2}, \cite{Bouchaud}
where the general relation 
between subdiffusive behaviour and diverging
waiting times has been treated in detail. 

\subsection{The process $X(t)=X(s(t))$}
       
We are interested in the properties of the variable X with respect
to the physical time t. Therefore, we have to consider the inverse of the
stochastic process $t=t(s)$:
\begin{equation}
s=t^{-1}(t)=s(t) \qquad .
\end{equation}
The stochastic process $X(s(t))$ then is described by the joint
probability distribution
\begin{equation}\label{ret}
f(x_2,t_2;x_1,t_1)=
<\delta(x_2-X(s_2))\delta(s_2-s(t_2))
\delta(x_1-X(s_1))\delta(s_1-s(t_1))> \qquad .
\end{equation}
The N-point distributions are determined in a similar way. Introducing
the probability distribution $h$ for the inverse process $s(t)$,
\begin{eqnarray}
h(s,t)&=&<\delta(s-s(t))> \qquad , \nonumber \\
h(s_2,t_2;s_1,t_1)&=&<\delta(s_2-s(t_2)) \delta(s_1-s(t_1))> \qquad ,
\end{eqnarray}
we can calculate the pdf of the process $X(t)=X(s(t))$ as a
function of the physical time by eliminating the path variables $s_i$:
\begin{eqnarray}
\label{Joint1}
f(x_2,t_2;x_1,t_1)&=&\int_0^\infty ds_1 \int_0^\infty ds_2\: h(s_2,t_2;s_1,t_1)
f_1(x_2,s_2;x_1,s_1) \qquad.
\end{eqnarray}
This relationship is due to the fact that the processes $X(s)$ and
$t(s)$ are statistically independent. In that case,
the expectation values in (\ref{ret}) factorize. Equation
(\ref{Joint1}) can be generalized to N times. In fact, one may 
turn over to a path integral representation:
\begin{equation}
f(x(t))= \int \mathcal{ D}s(t) h(s(t)) f_1(x(s(t)))\qquad.
\end{equation}
However, we do not investigate this path integral further.\\
\\
\begin{figure}
\centering
\includegraphics[height=6cm]{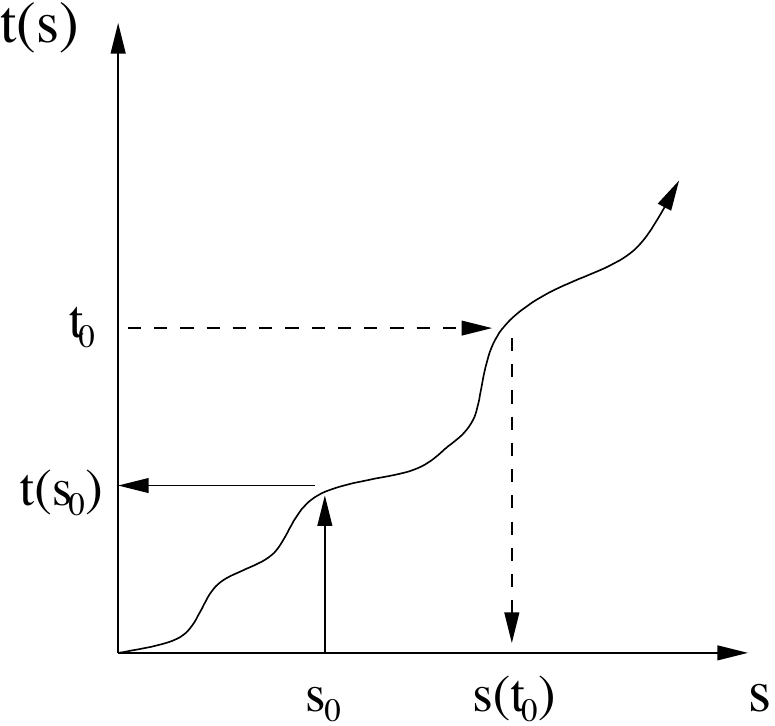}
\caption{Sketch of the process $t(s)$ which 
relates the arc length $s$ to
physical time $t$. Since the increment $\tau(s)$ of eq.(\ref{Langevin2}) is positive,
the curve $t(s)$ is monotonically increasing, implying the validity of
the relation (\ref{s-t-relation}).
.}
\end{figure}
The probability distribution $h$ can be determined with the help of
the cumulative distribution function of $s(t)$. Since the process $t(s)$ has the
property (for $s>0$) $s_2 > s_1 \rightarrow t(s_2) > t(s_1)$,
one has the relationship
\begin{eqnarray}
\label{s-t-relation}
\Theta(s-s(t))=1-\Theta(t-t(s))  \qquad.
\end{eqnarray}
Here, we have introduced the Heaviside step function:
$\Theta(x)=1$
for $x > 0$ and $\Theta(x)=0$ for $x < 0$, $\Theta(x=0)=1/2$.
The validity of eq.(\ref{s-t-relation}) becomes evident from an inspection of
fig. 1: The function $\Theta(s-s(t))$ equals one in the region above the curve
$t=t(s)$, whereas $\Theta(t-t(s)$ equals one in the region below
the
curve $t=t(s)$. On the curve
$\Theta(s-s(t))=1/2=\Theta(t-t(s))$.
\\
An immediate consequence is the following connection among the
cumulative distribution functions of the processes $t(s)$ and $s(t)$:
\begin{eqnarray}
\label{CDF}
<\Theta(s-s(t))>&=&1-<\Theta(t-t(s))>  \qquad, \nonumber \\
<\Theta(s_2-s(t_2))\Theta(s_1-s(t_1)>&=&<(1-\Theta(t_2-t(s_2)))(1-\Theta(t_1-t(s_1)))> \nonumber \\
&=&1-<\Theta(t_2-t(s_2))>-<\Theta(t_1-t(s_1))> \nonumber \\
&&+<\Theta(t_2-t(s_2))\Theta(t_1-t(s_1)> \qquad.
\end{eqnarray}
Simple differentiation of eq.(\ref{CDF}) yields the probability density function
$h$ of the process $s(t)$:
\begin{eqnarray}
\label{Main1}
h(s,t) &=&-\frac{\partial }{\partial s} <\Theta(t-t(s))> \qquad,
\nonumber \\
h(s_2,t_2;s_1,t_1) &=&
\frac{\partial}{\partial s_1}\frac{\partial}{\partial s_2}
<\Theta(t_2-t(s_2)) \Theta(t_1-t(s_1))> \qquad.
\end{eqnarray}
Furthermore, since for $t=0$ we have the correspondence $s=0$, the usual
boundary conditions hold:
\begin{eqnarray}
h(s,0) &=& \delta(s) \qquad, \nonumber \\
h(s_2,t_2;s_1,0) &=& h(s_2,t_2) \delta(s_1) \qquad, \nonumber \\
h(s_2,t_2\rightarrow t_1;s_1,t_1) &=& \delta(s_2-s_1) h(s_1,t_1) \qquad,
\end{eqnarray}
and can be verified from eq.(\ref{Main1}).

\section{Determination of the Probability distributions 
$p(s,t)$: L\'evy-stable processes}

In the following we shall consider the joint multiple times pdf 
of the L\'evy-stable process
(\ref{Langevin2})
of order $\alpha$.
Simple integration of (\ref{Langevin2}) yields
\begin{eqnarray}
\label{t_int}
t(s_i)=\int_0^{s_i} ds' \tau(s')\qquad,
\end{eqnarray}
where we assume $\tau(s) > 0$.
Additionally, we consider the characteristic function for $\omega=i
\lambda$. This defines the Laplace transform
\begin{equation}
Z(\lambda_2,s_2;\lambda_1,s_1):= \mathcal{L}\{p(t_2,s_2;t_1,s_1)\}=
\int_0^{\infty} dt_2\int_0^{\infty} dt_1 \: e^{-\lambda_2 t_2-\lambda_1 t_1}\: p(t_2,s_2;t_1,s_1)
\qquad.
\end{equation}
It will become clear below that working with Laplace transforms is more convenient for manipulating the pdf's of
process (\ref{Langevin2}) in the present context.

\subsection{One-sided L\'evy-stable processes: Single time}

At this point we have to introduce specific properties of 
the L\'evy-stable process. 
L\'evy distributions $L_{\alpha,\beta}(x)$ are defined by
two parameters \cite{Kolmogorov}, \cite{Yanovsky}:
$\alpha$ characterizes the asymptotic behaviour of 
the stable distribution for large x and hence the
critical order of diverging moments. $\beta$ characterizes the asymmetry.
In the present case $\tau > 0$ and
the distribution is maximally asymmetric $p(t<0,s)=0$. This leads to
$\beta=1$. In the following 
we denote the L\'evy distribution $L_{\alpha,\beta}(x)$
for $\beta=1$ by $L_\alpha(x)$.

Let us motivate the consideration of L\'evy statistics. To this end we
consider the characteristic function, which we write in the form:
\begin{equation}
Z(\lambda,s)=\:<e^{-\lambda s^{1/\alpha}\frac{1}{s^{1/\alpha}}
\int_0^s ds' \tau(s')}> \qquad ,
\end{equation}
where $\alpha$ is a certain parameter. The choice
$Z(\lambda,s)=\tilde Z(\lambda^\alpha s)$ leads to a scale invariant pdf
$p(t,s)=1/s^{1/\alpha} P(\frac{t}{s^{1/\alpha}})$ \cite{MeerPRE}.

As a result, the characteristic function takes the form
\begin{equation}
\label{CF_n1}
Z(\lambda,s)=e^{-\lambda^\alpha s} \qquad,
\end{equation}
where we assume $0 < \alpha <1$.\\
The probability distribution then becomes
\begin{equation}
p(t,s)=\frac{1}{s^{1/\alpha}}L_\alpha(\frac{t}{s^{1/\alpha}}) \qquad,
\end{equation}
where $L_\alpha(t)$ denotes the one sided L\'evy stable distribution whose
Laplace transform is $\mathcal{L}\{L_\alpha(t)\}=e^{-\lambda^\alpha}$.

\subsection{Multiple times}

The joint pdf of the Levy process $t(s)$ has been introduced in eq.(\ref{pdf2}). Starting with this
definition the derivation of the explicit expression for the pdf is straightforward
and clearly reveals the Markovian character of this process. The characteristic function
is given as Laplace transform of eq.(\ref{pdf2}):
\begin{eqnarray}
\label{CF_Def}
Z(\lambda_2,s_2;\lambda_1,s_1)
 &=& \int_0^{\infty}dt_2
\int_0^{\infty}dt_1 \: e^{-\lambda_2 t_2 -\lambda_1 t_1}\: p(t_2,s_2;t_1,s_1)
\nonumber \\
 &=&
<e^{-\lambda_2 \int_0^{s_2} ds' \tau(s')-\lambda_1
\int_0^{s_1} ds' \tau(s')}> \qquad.
\end{eqnarray}
For further evaluating this expression we have to distinguish between the cases $s_2>s_1$
and $s_1>s_2$. With a given ordering of $s_2,s_1$ we can rearrange the integrals and
write $Z$ as a sum of two contributions:
\begin{eqnarray}
Z(\lambda_2,s_2;\lambda_1,s_1)&=&\Theta(s_2-s_1)
<e^{-\lambda_2 \int_{s_1}^{s_2} ds' \tau(s')-(\lambda_1+\lambda_2)
 \int_0^{s_1} ds' \tau(s')}> \nonumber \\
 &&+\Theta(s_1-s_2)
<e^{-\lambda_1 \int_{s_2}^{s_2} ds' \tau(s')-(\lambda_1+\lambda_2)
 \int_0^{s_2} ds' \tau(s')}>
\qquad.
\end{eqnarray}
Here the expectation values factorize due to statistical independence of the increments $\tau$
and can be expressed according to eq.(\ref{CF_n1}):
\begin{eqnarray}
\label{Main2}
Z(\lambda_2,s_2;\lambda_1,s_1)&=&\Theta(s_2-s_1)
e^{-s_1(\lambda_1+\lambda_2)^\alpha}e^{-(s_2-s_1)
\lambda_2^\alpha} \nonumber \\
&&+\Theta(s_1-s_2)e^{-s_2(\lambda_1+\lambda_2)^\alpha}e^{-(s_1-s_2)
\lambda_1^\alpha} \qquad.
\end{eqnarray}
This is the characteristic function of the Levy process for multiple times. The appearance
of the exponents $(\lambda_1+\lambda_2)^\alpha$ is characteristic in this context and carries over
to the pdf of the inverse process.
We obtain the pdf $p(s_2,t_2;s_1,t_1)$ after performing the
inverse Laplace transform of
eq.(\ref{Main2}). The result is
\begin{eqnarray}
p(t_2,s_2;t_1,s_1)&=&\Theta(s_2-s_1)\frac{1}{(s_2-s_1)^{1/\alpha}}
L_\alpha\left(\frac{t_2-t_1}{(s_2-s_1)^{1/\alpha}}\right) \frac{1}{s_1^{1/\alpha}}
L_\alpha\left(\frac{t_1}{s_1^{1/\alpha}}\right) \nonumber \\
&&+\Theta(s_1-s_2)\frac{1}{(s_1-s_2)^{1/\alpha}}
L_\alpha\left(\frac{t_1-t_2}{(s_1-s_2)^{1/\alpha}}\right) \frac{1}{s_2^{1/\alpha}}
L_\alpha\left(\frac{t_2}{s_2^{1/\alpha}}\right) \qquad.
\end{eqnarray}
This expression explicitly exhibits the Markovian nature of the
process. The conditional pdf $p(t_2,s_2|t_1,s_1)$ for $s_2>s_1$ is just:
\begin{equation}
p(t_2,s_2|t_1,s_1)= \frac{1}{(s_2-s_1)^{1/\alpha}}
L_\alpha\left(\frac{t_2-t_1}{(s_2-s_1)^{1/\alpha}}\right) \qquad.
\end{equation}
We remind the reader that $L_\alpha(x)=0$ for negative values
of $x$.
The expression for the joint pdf for multiple points is obvious.

\section{The Probability Distributions $h(s,t)$}

The pdf's $h(s,t)$, $h(s_2,t_2;s_1,t_1)$
of the inverse process $s=s(t)$ can be obtained from the pdf's of the
process $t=t(s)$
with the help of relationship eq.(\ref{Main1}). We shall
consider the  single- and multiple-time cases separately. Again, due to the
simple form of the Levy distributions in Laplace space,
we perform most of the calculations with Laplace transforms.

\subsection{Single time}

Using the notation $\tilde{h}(s,\lambda)=\mathcal{L}\{h(s,t)\}$ for
the Laplace transform of $h(s,t)$ with respect to t, the
relation eq.(\ref{Main1}) reads:
\begin{eqnarray}\label{ph_L_n1}
\tilde{h}(s,\lambda)
&=&-\frac{\partial}{\partial s}<\frac{1}{\lambda}e^{-\lambda t(s)}>\:
=-\frac{\partial}{\partial s}\:\frac{1}{\lambda}\:Z(s,\lambda)\qquad.
\end{eqnarray}

The derivative with respect to $s$ is easily performed with eq.(\ref{CF_n1}) and
leads to the solution $\tilde{h}(s,\lambda)$:
\begin{eqnarray}
\label{Barkai_L}
\tilde{h}(s,\lambda)&=&\lambda^{\alpha-1}e^{-s\lambda^\alpha}\qquad.
\end{eqnarray}
This expression has already been derived in \cite{Fogedby} --- however without
giving a `simple physical argument'. Here the
derivation is clearly based on eq.(\ref{s-t-relation}) 
which relates the L\'evy-stable
process and its inverse.\\
The inverse Laplace transform of eq.(\ref{Barkai_L})
is known and has been calculated in \cite{Barkai}:
\begin{equation}
\label{Barkai_n}
h(s,t)=\frac{1}{\alpha} \frac{t}{s^{1+1/\alpha}}
L_\alpha(\frac{t}{s^{1/\alpha}}) \qquad.
\end{equation}
Moreover, in \cite{Bingham} the single time distribution $h(s,t)$ has been
identified as the Mittag-Leffler distribution:
\begin{equation}
h(s,t)= \sum_{n=0}^{\infty} \frac{(-st^\alpha)^n}{\Gamma(1+n\alpha)}\qquad.
\end{equation}

Here we have obtained the pdf of $s(t)$ for single times.
Therefore, a complete characterization of the inverse process is given in this case.\\
However in order to derive an evolution equation for the pdf of the
process $X(s(t))$ we require an equation which determines $h(s,t)$.

From eq.(\ref{Barkai_L}) it is evident that $\tilde{h}(s,\lambda)$ obeys
the differential equation
\begin{eqnarray}
\label{Main3_L_n1}
-\frac{\partial}{\partial s}\tilde{h}(s,\lambda) = \lambda^\alpha \tilde{h}(s,\lambda)
\end{eqnarray}
with the initial condition $\tilde{h}(0,\lambda)=\lambda^{\alpha-1}$
for $s=0$.
Hence, Laplace inversion yields a fractional evolution equation for $h(s,t)$:
\begin{eqnarray}
\label{Main3_n1}
\frac{\partial}{\partial t}h(s,t) &=&
-_0D_{t}^{1-\alpha}\frac{\partial}{\partial s} h(s,t) \qquad .
\end{eqnarray}
The operator $_0D_{t}^{1-\alpha}$ denotes the Riemann-Liouville fractional
differential operator, a possible generalization of integer 
order differentiation
and integration to fractional orders (see Appendix B). 
For a discussion of fractional derivatives
we refer the reader to \cite{Podlubny}.

\subsection{Multiple times}

The statistical characterization of the process $s(t)$ for multiple
times has been investigated from a mathematical point of view in the
work of Bingham \cite{Bingham} already in 1971. 
He derived the following relationships for the moments $<s(t_N)...s(t_1)>$:
\begin{equation}\label{Bing}
\frac{\partial^N }{\partial t_1...\partial t_N} 
<s(t_N)...s(t_1)>\:= \frac{1}{\Gamma(\alpha)^N}[t_1(t_2-t_1)...(t_N-t_{N-1})]^{\alpha-1}
\end{equation}
This equation can be obtained from the previous
relation (\ref{Main1}), which inferes the following
relationship between the probability densities $p(t,s)$ and $h(s,t)$:
\begin{eqnarray}
\frac{\partial }{\partial t} h(s,t) &=& -\frac{\partial }{\partial
s}p(t,s)
\nonumber \\
\frac{\partial^2 }{\partial t_1\partial t_2} h(s_2,t_2;s_1,t_2)
&=& \frac{\partial^2 }{\partial
s_2\partial
s_1}p(t_2,s_2;t_1,s_1)
\nonumber \\
\frac{\partial^N }{\partial t_1...\partial t_N} h(s_N,t_N;...;s_1,t_2)
&=& (-1)^N\frac{\partial^N }{\partial
s_N...\partial
s_1}p(t_N,s_N;...;t_1,s_1)\qquad.
\end{eqnarray}

In the following we shall derive explicit expressions for these
moments and show that instead of (\ref{Bing}) fractional equations can
be used for their determination. 
Based on eq.(\ref{Main1}) and eq.(\ref{Main2})
the derivation of an expression for the Laplace transform
$\tilde{h}(s_2,\lambda_2;s_1,\lambda_1):=\mathcal{L}\{h(s_2,t_2;s_1,t_1)\}$
is obtained in a way analogous to the single-time case.\\
\\
We start by considering eq.(\ref{Main1}) in Laplace-space:
\begin{eqnarray}
\label{Main1_L}
\tilde{h}(s_2,\lambda_2;s_1,\lambda_1)
&=&\frac{\partial}{\partial s_1}\frac{\partial}{\partial s_2}
<\frac{1}{\lambda_2}e^{-\lambda_2 t(s_2)}\frac{1}{\lambda_1}e^{-\lambda_1 t(s_1)}> \nonumber \\
&=&\frac{\partial}{\partial s_1}\frac{\partial}{\partial s_2}\: \frac{1}{\lambda_1 \lambda_2}
\: Z(\lambda_2,s_2;\lambda_1,s_1) \qquad.
\end{eqnarray}
Using eq.(\ref{Main2}) we can perform the derivatives of
$Z(\lambda_2,s_2;\lambda_1,s_1)$ with respect to $s_1$, $s_2$:
\begin{eqnarray}
\label{EqH_L}
\tilde{h}(s_2,\lambda_2;s_1,\lambda_1)&=&\delta(s_2-s_1)\frac{\lambda_1^\alpha-(\lambda_1+\lambda_2)^\alpha
+\lambda_2^\alpha}{\lambda_1 \lambda_2}e^{-s_1(\lambda_1+\lambda_2)^\alpha} \nonumber \\
&&+ \Theta(s_2-s_1)\frac{(\lambda_2^\alpha)
((\lambda_1+\lambda_2)^\alpha-\lambda_2^\alpha)}
{\lambda_1 \lambda_2}
e^{-(\lambda_1+\lambda_2)^\alpha s_1}
e^{-\lambda_2^\alpha (s_2-s_1)}\nonumber \\
&&+ \Theta(s_1-s_2)\frac{(\lambda_1^\alpha)
((\lambda_1+\lambda_2)^\alpha-\lambda_1^\alpha)}
{\lambda_1 \lambda_2}
e^{-(\lambda_1+\lambda_2)^\alpha s_2}
e^{-\lambda_1^\alpha (s_1-s_2)}
\qquad.
\end{eqnarray}

As a result we have obtained the Laplace transform of the joint pdf
$h(s_2,t_2;s_1,t_1)$. Unfortunately,
a closed form of the inverse Laplace transform could not be calculated. The given solution
$\tilde{h}$ can be readily used however to derive meaningful expressions which characterize
the inverse process $s(t)$.

\subsubsection{Moments of the inverse process}

In order to obtain further information about the process $s(t)$
for multiple times we
calculate the moments of the pdf. Let us first demonstrate how
this can be achieved
for the simple case $<s(t_1)s(t_2)>$. This moment is defined from the pdf $h(s_2,t_2;s_1,t_1)$
as:
\begin{eqnarray}
<s(t_1)s(t_2)>&=&\int_0^\infty ds_1 \:\int_0^\infty ds_2\:s_1 s_2
\:h(s_2,t_2;s_1,t_1) \nonumber \\ &=&\mathcal{L}^{-1}\left\{\int_0^\infty ds_1
\:\int_0^\infty ds_2\:s_1 s_2\:
\tilde{h}(s_2,\lambda_2;s_1,\lambda_1)\right\}\qquad, \end{eqnarray} where the
last step follows by interchanging inverse Laplace transform  and integration.
The integrations with respect to $s_1,s_2$ can be simply performed with the help of expression eq.(\ref{Main1_L}).
The result is:
\begin{eqnarray}
\label{Moment_L}
\int_0^\infty ds_1 \:\int_0^\infty ds_2\:s_1 s_2
\:\tilde{h}(s_2,\lambda_2;s_1,\lambda_1)=
(\lambda_1+\lambda_2)^{-\alpha}\left\{\frac{\lambda_1^{-\alpha-1}}
{\lambda_2}+\frac{\lambda_2^{-\alpha-1}} {\lambda_1}\right\} \qquad.
\end{eqnarray}
Now the inverse Laplace transform leads to an analytical solution for $<s(t_1)s(t_2)>$ (see Appendix B):
\begin{eqnarray}
\label{Moment}
<s(t_1)s(t_2)>\:&=&
\Theta(t_2-t_1)\left\{\frac{1}{\Gamma(2\alpha+1)}t_1^{2\alpha}+\frac{1}{\Gamma
(\alpha+1)^2}\:t_1^\alpha
t_2^\alpha\:F\left(\alpha,-\alpha;\alpha+1;\frac{t_1}{t_2}\right)\right\}
\nonumber \\ &&+\Theta(t_1-t_2)\left\{\frac{1}{\Gamma(2\alpha+1)}t_2^{2\alpha}+
\frac{1}{\Gamma(\alpha+1)^2}\:t_1^\alpha
t_2^\alpha\:F\left(\alpha,-\alpha; \alpha+1;\frac{t_2}{t_1}\right)\right\}.
\end{eqnarray}
Here $F(a,b;c;z)$ denotes the hypergeometric function (see e.g.
Ch.15 in \cite{Abram}).\\

One notices that in the limit
$t_2 \rightarrow t_1$ expression (\ref{Moment})
agrees with the second moment $<s(t)^2>$:
\begin{eqnarray}
<s(t)^2>&=&
\mathcal{L}^{-1}\left\{\int_0^\infty s^2\lambda^{\alpha-1}e^{-s\lambda^\alpha}ds\right\}
=\frac{2}{\Gamma(2\alpha+1)}t^{2\alpha}\qquad,
\end{eqnarray}
where eq.(\ref{Barkai_L}) has been used. The simple single time moment
$<s(t)>$ is given as
$<s(t)>\:=\mathcal{L}^{-1}\left\{\lambda^{-\alpha-1}\right\}
=\frac{1}{\Gamma(\alpha+1)}t^\alpha$.\\ \\
The calculation of higher order moments essentially
follows the same steps.\\
Furthermore, we introduce the
operator $ \left(\frac{\partial}{\partial t_1} + \frac{\partial}{\partial t_2}
\right)^{1-\alpha}$
in the sense of the single-time Riemann-Liouville fractional
differential operator:
$\mathcal{L}\{\left(\frac{\partial}{\partial t_1} + \frac{\partial}
{\partial
t_2}\right)^{-\alpha}g(t_1,t_2)\}=(\lambda_1+\lambda_2)^{-\alpha}\tilde{g}
(\lambda_1,\lambda_2)$ (see Appendix A). An explicit expression in terms of an
integral reads: \begin{equation}
\left(\frac{\partial }{\partial t_1}+\frac{\partial }{\partial
t_2}\right)^{-\alpha}
g(t_1,t_2)=\frac{1}{\Gamma(\alpha)}
\int_0^{Min(t_1,t_2)} dt'\:t'^{\alpha-1} g(t_1-t',t_2-t')
\qquad .
\end{equation}

Using this fractional differential operator, we are in the position
to write down a simple recursion relation for arbitrary moments of
$h(\{s_i,t_i\})$. The second moment eq.(\ref{Moment_L}) reads:
\begin{eqnarray}
\label{Moment2}
<s(t_1)s(t_2)>\:=\left(\frac{\partial}{\partial t_1}+\frac{\partial}{\partial t_2}\right)^{-\alpha}
\left\{<s(t_1)>+<s(t_2)>\right\}
\qquad.
\end{eqnarray}
This immediately leads to (we assume $t_2 > t_1$):
\begin{equation}
<s(t_2)s(t_1)>\:=\left [ _0D_{t_1}^{-\alpha} \lbrace <s(t_2-\tilde
t_1+t_1)>+<s(t_1)>
\rbrace \right]_{\tilde t_1=t_1} \qquad .
\end{equation}
The explicit expression allows one to obtain the fusion rule
\begin{eqnarray}
\lim_{t_2\rightarrow t_1} <s(t_2)s(t_1)>\:=\:<s(t_1)^2>
= 2 \frac{1}{\Gamma(\alpha)}
\int_0^{t_1} dt'\:t'^{\alpha-1} <s(t_1-t')>\:= 2 _0D_{t_1}^{-\alpha} s(t_1).
\end{eqnarray}

The calculation of the third order moment $<s(t_1)s(t_2)s(t_3)>$ along
the same lines yields the result:
\begin{eqnarray}\label{drei}
<s(t_1)s(t_2)s(t_3)>\:=\left(\frac{\partial}{\partial t_1}+\frac{\partial}{\partial t_2}
+\frac{\partial}{\partial t_3}\right)^{-\alpha}
&\{&<s(t_1)s(t_2)>+<s(t_1)s(t_3)> \nonumber \\
&&+<s(t_2)s(t_3)>\}
\qquad.
\end{eqnarray}
The third moment is obtained via fractional integration of the sum of second order moments. In the general case,
the n-th order moment is calculated by fractional integration with respect to n times of the sum
of all permutations of $n-1$ order moments.\\
Due to the representation of the fractional operator
\begin{equation}
\label{three_reps}
\left(\frac{\partial }{\partial t_1}+\frac{\partial }{\partial
t_2}+\frac{\partial }{\partial t_3}\right)^{-\alpha} g(t_1,t_2,t_3)
= \frac{1}{\Gamma(\alpha)} \int_0^{Min(t_1,t_2,t_3)}
dt'\:t'^{\alpha-1} g(t_1-t',t_2-t',t_3-t'),
\end{equation}
we can derive the fusion rule
\begin{eqnarray}
\lim_{t_3 \rightarrow t_1+0} <s(t_3)s(t_2)s(t_1)>&=& \frac{1}{\Gamma(\alpha)} \int_0^{t_1} dt'\:t'^{\alpha-1}
\lbrace <s(t_1-t') s(t_1-t')>+
2 <s(t_2-t') s(t_1-t')> \rbrace
\nonumber \\
&=& _0D_{t_1}^{-\alpha} \lbrace <s(t_1)s(t_1)>+2
<s(t_2-\tilde t_1+t_1)s(t_1)>
\rbrace_{\tilde t_1=t_1}\qquad.
\end{eqnarray}

The fusion $t_2 \rightarrow t_1$ leads to
\begin{equation}
<s(t_1)^3>\:= 3 _0D_{t_1}^{-\alpha} <s(t_1)^2>\:= 6 D_{t_1}^{-\alpha}
D_{t_1}^{-\alpha}
<s(t_1)>\:= 6 _0D_{t_1}^{-2\alpha} <s(t_1)>.
\end{equation}

The n-th order generalization reads:
\begin{equation}
<s(t)^n>\:= n! ~ _0D_t^{-(n-1)\alpha} <s(t)> \qquad.
\end{equation}

This equation can also be derived directly from $\tilde{h}(s,\lambda)$.
Thus one can obtain a complete characterization of the process
$s(t)$ based on eq.(\ref{EqH_L}) or eq.(\ref{Main1_L}) respectively. Below, we
shall show how to obtain these results on the basis of an evolution equation for
the multipoint pdf $h(s_1,t_1;...;s_N,t_N)$.

\subsubsection{The structure of the N-time pdf}

From eq.(\ref{Main1}) one can derive the general form of the pdf $h$ of the inverse process $s(t)$.
The two times pdf reads (here we assume the case $s_2>s_1$ for simplicity)
\begin{eqnarray}
h(s_2,t_2;s_1,t_1)&=&\frac{\partial}{\partial s_1}
\frac{\partial}{\partial s_2}\int_0^{t_1}dt_1'\int_0^{t_2}dt_2'
\:p(t_2'-t_1',s_2-s_1)\:p(t_1',s_1) \nonumber \\
&=&-\frac{\partial}{\partial
s_1}\int_0^{t_1}dt_1'\:h(s_2-s_1,t_2-t_1')\:p(t_1',s_1) \qquad.
\end{eqnarray}

We define
\begin{equation}\label{zwei}
H(s_2-s_1,t_2-t_1;s_1-s_0,t_1-t_0)=
-\frac{\partial }{\partial s_1} \int_0^{t_1} dt'_1 \:h(s_2-s_1,t_2-t_1')\:
p(t_1'-t_0,s_1-s_0).
\end{equation}

The form of the three times pdf is obtained in the same way and reads for $s_3>s_2>s_1$:
\begin{eqnarray}
h(s_3,t_3;s_2,t_2;s_1,t_1)&=&\frac{\partial}{\partial
s_1}\frac{\partial}{\partial s_2}\int_0^{t_1}dt_1'\int_0^{t_2}dt_2'\:
h(s_3-s_2,t_3-t_2')\:p(t_2'-t_1',s_2-s_1)\:p(t_1',s_1) \nonumber\\
\end{eqnarray}
with a straightforward extension to the general case.\\
With the help of eq.(\ref{zwei}) this expression can be represented according to
\begin{equation}
h(s_3,t_3;s_2,t_2;s_1,t_1)=
-\frac{\partial }{\partial s_1} \int_0^{t_1} dt_1'\:
H(s_3-s_2,t_3-t_2;s_2-s_1,t_2-t_1')\:p(t_1',s_1)\qquad.
\end{equation}
Recursively, we may define higher order functions
\begin{eqnarray}
&H^N&(s_N-s_{N-1},t_N-t_{N-1};...;t_1-t_0,s_1-s_0)
\nonumber \\
&=& -\frac{\partial }{\partial s_1} \int_0^{t_1} dt_1' \:H^{N-1}(
s_N-s_{N-1},t_N-t_{N-1};...;s_2-s_1,t_2-t_1')\:p(t_1'-t_0,s_1,s_0).
\end{eqnarray}
The integrals cannot simply be evaluated and
the relations are formal. However, they show the underlying
mathematical structure of the statistical description of
the inverse process $s(t)$.

\subsubsection{Fractional evolution equation}

In analogy to the single time case, where we have specified a
fractional differential equation for $h(s,t)$, we now
establish an evolution equation
for $h(s_2,t_2;s_1,t_1)$.\\
From eq.(\ref{EqH_L}) it is evident that the following equation holds:
\begin{eqnarray}
\label{Main3_L}
\left(\frac{\partial}{\partial s_1}+\frac{\partial}{\partial s_2}\right)
\tilde{h}(s_2,\lambda_2;s_1,\lambda_1)
&=& -(\lambda_1+\lambda_2)^\alpha \tilde{h}(s_2,\lambda_2;s_1,\lambda_1)\qquad
\end{eqnarray}
with initial conditions
\begin{eqnarray}
\label{initial_L}
\tilde{h}(0,\lambda_2;0,\lambda_1)&=&\frac{\lambda_1^\alpha-
(\lambda_1+\lambda_2)^\alpha+\lambda_2^\alpha}
{\lambda_1 \lambda_2} \qquad, \nonumber \\
\tilde{h}(s_2,\lambda_2;0,\lambda_1)&=&\frac{(\lambda_2^\alpha)
((\lambda_1+\lambda_2)^\alpha-\lambda_2^\alpha)}
{\lambda_1 \lambda_2}e^{-\lambda_2^\alpha s_2} \qquad, \nonumber \\
\tilde{h}(0,\lambda_2;s_1,\lambda_1)&=&\frac{(\lambda_1^\alpha)
((\lambda_1+\lambda_2)^\alpha-\lambda_1^\alpha)}
{\lambda_1 \lambda_2}
e^{-\lambda_1^\alpha s_1}
\qquad.
\end{eqnarray}
A common way to solve first order partial differential equations is the method of characteristics.
Applying this method to eq.(\ref{Main3_L}) with the given initial condition
for each case ,
one obtains the correct expressions eq.(\ref{EqH_L}). Therefore eq.(\ref{Main3_L})
determines the pdf in Laplace space.\\
Consequently, upon performing the inverse Laplace transform,
we derive that $ h(s_2,t_2;s_1,t_1)$
obeys the fractional evolution equation
\begin{eqnarray}
\label{Main3}
\left(\frac{\partial}{\partial t_1}
+\frac{\partial}{\partial t_2}\right)h(s_2,t_2;s_1,t_1)
&=& -\left(\frac{\partial}{\partial t_1}
+\frac{\partial}{\partial t_2}\right)^{1-\alpha}
\left(\frac{\partial}{\partial s_1}+\frac{\partial}{\partial s_2}
\right) h(s_2,t_2;s_1,t_1) \qquad,
\end{eqnarray}
where the fractional differential operator $\left(\frac{\partial}{\partial t_1}
+\frac{\partial}{\partial t_2}\right)^{1-\alpha}$ has been defined according to
$\left(\frac{\partial}{\partial t_1}
+\frac{\partial}{\partial
t_2}\right)^{1-\alpha}F(t_2,t_1):=\left(\frac{\partial}{\partial t_1}
+\frac{\partial}{\partial t_2}\right)\left(\frac{\partial}{\partial t_1}
+\frac{\partial}{\partial t_2}\right)^{-\alpha}F(t_2,t_1)$. The appearance of
fractional time derivatives in eq.(\ref{Main3}) reveals the non-Markovian
character of the stochastic process $s(t)$ and as a consequence of the coupled
process $X(s(t))$.\\
\\
The extension of the above result to n times is straightforward:
\begin{eqnarray}
\label{Main3_n}
\left( \sum_{i=1}^N\frac{\partial}{\partial t_i} \right)
h(\{s_i,t_i\}) &=&
-\left( \sum_{i=1}^N\frac{\partial}{\partial t_i} \right)^{1-\alpha}
\left( \sum_{i=1}^N \frac{\partial}{\partial s_i} \right)
h(\{s_i,t_i\}) \qquad.
\end{eqnarray}
Again we want to emphasize that this single evolution equation with the proper
initial condition sufficiently assets the pdf for multiple times.\\
The above equation may also be used to calculate the moments
$<s(t_N)...s(t_1)>$, which already have been specified above.
The fractional evolution equation (\ref{Main3_n}) inferes the
following relationship among the moments $<s(t_N)...s(t_1)>$:
\begin{eqnarray}
\left( \sum_{i=1}^N\frac{\partial}{\partial t_i} \right)
<s(t_N)...s(t_1)> &=&
\left( \sum_{i=1}^N\frac{\partial}{\partial t_i} \right)^{1-\alpha}
\lbrace <s(t_{N-1})...s(t_1)>+Permut \rbrace
 \qquad.
\end{eqnarray}
These equations are equivalent to the chain of equations (\ref{drei})
obtained by a direct inspection of the pdf's.

\section{Two-Time Moments of the Diffusion process}

In this last section we focus on the usual diffusion process, i.e.
we consider the Fokker-Planck operator
\begin{equation}
L= \frac{\partial^2}{\partial x^2} \qquad.
\end{equation}
In this case, the moments are polynomials in $s$ and we may directly use
the results of the preceding session:
\begin{equation}
<x(s_2)x(s_1)>\:=\Theta(s_2-s_1)s_1 +\Theta(s_1-s_2)s_2 \qquad.
\end{equation}
The corresponding moment with respect to time t is given by
\begin{equation}
<x(t_2)x(t_1)>\:= \int_0^\infty \int_0^\infty ds_1 ds_2
\:h(s_2,t_2;s_1,t_1)<x(s_2)x(s_1)>\qquad.
\end{equation}
The integrations can be performed by inserting the pdf
$h$ in Laplace space:
\begin{equation}
\mathcal{L}\{< x(t_2) x(t_1)>\}\:=\:
\frac{(\lambda_1+\lambda_2)^\alpha}{\lambda_1 \lambda_2} \int_0^\infty
ds~ s \:e^{-(\lambda_1+\lambda_2)^\alpha s}
=\frac{1}{(\lambda_1+\lambda_2)^\alpha \lambda_1 \lambda_2} \qquad.
\end{equation}
The inverse transform leads to the result
\begin{eqnarray}
<x(t_2) x(t_1) >\:&=&\frac{1}{\Gamma(\alpha+1)} \lbrace\Theta(t_2-t_1)
t_1^\alpha + \Theta(t_1-t_2) t_2^\alpha \rbrace \nonumber \\
&=& \Theta(t_2-t_1)<s(t_1)>+\Theta(t_1-t_2)<s(t_2)> \qquad.
\end{eqnarray}

Similarly, we may calculate the moment $<x(t_2)^2x(t_1)^2>$:
\begin{equation}
<x(s_2)^2 x(s_1)^2>\:= s_2 s_1+2 \Theta(s_2-s_1) s_1^2+2\Theta(s_1-s_2)s_2^2
\qquad. \end{equation}
This yields
\begin{eqnarray}
<x(t_2)^2 x(t_1)^2>\:=\:<s(t_2) s(t_1)>+2 \Theta(t_2-t_1) <s(t_1)^2>
+2\Theta(t_1-t_2)
<s(t_2)^2>.
\end{eqnarray}
For the evaluation of $<x(s_2)^{2m}x(s_1)^{2n}>$ we may use
the properties of the moments of Gaussian processes
which read for $n>m$:
\begin{equation}
<x(s_2)^{2m}x(s_1)^{2n}>\:= A s_2^m s_1^n +B \Theta(s_2-s_1) s_1^{n-m}
s_2^m+B
\Theta(s_1-s_2) s_2^{n-m} s_1^m \qquad.
\end{equation}
The coefficients A, B, C can be evaluated by an application of Wick's theorem
for Gaussian processes.

The corresponding expression for the process $X(t)$ becomes accordingly:
\begin{eqnarray}
<x(t_2)^{2m}x(t_1)^{2n}> &=& A <s(t_2)^m s(t_1)^n> +B\Theta(t_2-t_1) <s(t_1)^{n-m}
s(t_2)^m>
\nonumber \\
&&+ B
\Theta(t_1-t_2) <s(t_2)^{n-m} s(t_1)^m> \qquad.
\end{eqnarray}
The calculation of the expectation values $<s(t_2)^{2m}s(t_1)^{2n}>$
has been discussed above.

\section{Conclusion}

Up to now the discussion of continuous time random walks and the
corresponding fractional
kinetic equations has been focused on single time probability
distributions only. On the basis of this pdf scaling behaviour of
moments have been compared with experiments. However, more information
has to be used in order to
assign a definite stochastic process to a non-Markovian process.
To this end we have considered multiple
times pdf for a certain class of stochastic processes.\\
Our approach
is based on the framework of
coupled Langevin equations (\ref{Langevin1}),(\ref{Langevin2})
devised by Fogedby as a realization of a continuous time random walk.
Here,
the solution for the N-times pdf's are given as an integral
transform of the pdf's of an accompanying Markovian
process.
We have shown that the non-Markovian character of this process can be traced
back to the properties of the inverse L\'evy-stable process.\\
The next step would be to compare these theoretical predictions with
the behaviour of physical systems which reveal subdiffusive behaviour. To
our knowledge multiple time statistics of such systems have not yet been
investigated experimentally. This would be of considerable interest.
We may expect that in some cases the consideration of multiple time
statistics may lead to a more precise characterization of the
underlying stochastic process.\\
\\
It is well-known, that for the single time case a fractional diffusion
equation can be derived, which determines the pdf $f(x,t)$,
\begin{equation}
f(x,t)=\int_0^\infty ds \:h(s,t) f_1(x,s) \qquad ,
\end{equation}
 as a
solution of
\begin{equation}
\frac{\partial }{\partial t} f(x,t)=\:_0D_t^{1-\alpha} L_{FP}f(x,t)
\qquad .
\end{equation}
We would like to mention that a similar equation can be derived for
the multiple times pdf $f(x_2,t_2;x_1,t_1)$. This will be discussed in
a future publication.  The present article is a starting point for the
investigation of multiple times pdf's of the coupled Langevin
equations of Fogedby.

\acknowledgements
We gratefully acknowledge support by the Deutsche
Forschungsgemeinschaft and wish to thank
R. Hillerbrand, O. Kamps and T. D. Frank for helpful discussions.

\appendix

\section{Fractional differential operator}

The Riemann-Liouville fractional integral is defined as a generalization of the
Cauchy formula to real orders $\alpha$:
\begin{eqnarray}
\label{frac_int}
_0D_t^{-\alpha}g(t)&:=&\frac{1}{\Gamma(\alpha)}\int_0^t
\frac{g(t')}{(t-t')^{1-\alpha}}\:dt' \nonumber \\
&=&\frac{1}{\Gamma(\alpha)}t^{\alpha-1}*g(t) \qquad.
\end{eqnarray}
Here $*$ denotes a Laplace convolution. Consequently performing the Laplace transformation is
straightforward and yields the well-known result:
\begin{eqnarray}
\mathcal{L}\{_0D_t^{-\alpha}g(t)\}=\lambda^{-\alpha}\tilde{g}(\lambda) \qquad.
\end{eqnarray}
>From eq.(\ref{frac_int}) the Riemann-Liouville fractional differential operator is obtained
by simple partial derivation:
\begin{eqnarray}
\label{frac_diff}
_0D_t^{1-\alpha}g(t):=\frac{\partial}{\partial t}\:_0D_t^{-\alpha}g(t) \qquad.
\end{eqnarray}

The extension of the fractional differential operator to two times $t_1$,$t_2$ is now obtained
in a way analogous to the steps above.\\
First we define the fractional integral operator of two times in Laplace-space:
\begin{eqnarray}
\mathcal{L}\left\{\left(\frac{\partial}{\partial t_1} + \frac{\partial}
{\partial
t_2}\right)^{-\alpha}g(t_1,t_2)\right\}:=(\lambda_1+\lambda_2)^{-\alpha}
\tilde{g} (\lambda_1,\lambda_2) \qquad. \end{eqnarray}

Furthermore the following equation holds:
\begin{eqnarray}
\int_0^\infty dt_1\int_0^\infty dt_2 \: e^{-\lambda_1t_1-\lambda_2t_2}\:
\frac{1}{\Gamma(\alpha)}t_1^{\alpha-1}\delta(t_2-t_1)&=&\int_0^\infty dt_1\:
e^{-t_1(\lambda_1+\lambda_2)}\frac{1}{\Gamma(\alpha)}t_1^{\alpha-1} \nonumber \\
&=&(\lambda_1+\lambda_2)^{-\alpha} \qquad.
\end{eqnarray}

In physical time the fractional integral operator can thus be considered
as an expression containing a two-fold Laplace convolution
with respect to $t_1$ and $t_2$, denoted with $**$:
\begin{eqnarray}
\left(\frac{\partial}{\partial t_1} + \frac{\partial}
{\partial t_2}\right)^{-\alpha}g(t_1,t_2)&=&
\frac{1}{\Gamma(\alpha)}t_1^{\alpha-1}\delta(t_2-t_1)**g(t_2,t_1) \nonumber \\
&=&\frac{1}{\Gamma(\alpha)}\int_0^{t_1}dt_1'\int_0^{t_2}dt_2'\:t_1'^{\alpha-1}\delta(t_2'-t_1')\:
\:g(t_2-t_2',t_1-t_1') \qquad .
\end{eqnarray}
Here we can distinguish between the cases $t_2<t_1$ and $t_2>t_1$ which results
in eq.(\ref{three_reps}) The fractional differential operator of two times is
then corresponding to eq.(\ref{frac_diff}):
\begin{eqnarray}
\left(\frac{\partial}{\partial t_1} + \frac{\partial}
{\partial t_2}\right)^{1-\alpha}g(t_1,t_2):=
\left(\frac{\partial}{\partial t_1} + \frac{\partial}
{\partial t_2}\right)\left(\frac{\partial}{\partial t_1} + \frac{\partial}
{\partial t_2}\right)^{-\alpha}g(t_1,t_2) \qquad.
\end{eqnarray}

In the general N-times case the fractional integral operator takes the form
of an N-fold convolution
\begin{eqnarray}
\left(\sum_{i=1}^N\frac{\partial}{\partial t_i}\right)^{-\alpha}g(t_1,...,t_N)=
\frac{1}{\Gamma(\alpha)}t_1^{\alpha-1}\delta(t_N-t_{N-1})...\delta(t_2-t_1)
*...*g(t_1,...,t_N) \qquad,
\end{eqnarray}

with Laplace-transform
\begin{eqnarray}
\mathcal{L}\left\{\left(\sum_{i=1}^N\frac{\partial}{\partial t_i}\right)^{-\alpha}
g(t_1,...,t_N)\right\}=\left(\sum_{i=1}^N \lambda_i \right)^{-\alpha}\tilde{g}
(\lambda_1,...,\lambda_N) \qquad.
\end{eqnarray}

\section{Calculation of moments}

Using the results of the previous section we can explicitly write the second
order moment eq.(\ref{Moment2}) as convolution integrals:
\begin{eqnarray}
<s(t_1)s(t_2)>\:=
\frac{1}{\Gamma(\alpha)}\int_0^{t_1}dt_1'\int_0^{t_2}dt_2'\:t_1'^{\alpha-1}\delta(t_2'-t_1')\:
\:&\{&\frac{1}{\Gamma(\alpha+1)}(t_1-t_1')^{\alpha} +\frac{1}{\Gamma(\alpha+1)}(t_2-t_2')^{\alpha}\} \qquad.
\end{eqnarray}

If we distinguish between the cases $t_2>t_1$ and $t_1>t_2$ in order to perform
the integrations, we obtain:
\begin{eqnarray}
<s(t_1)s(t_2)>\:&=&\Theta(t_2-t_1)\left\{\frac{1}{\Gamma(2\alpha+1)}
t_1^{2\alpha}+\frac{1}{\Gamma(\alpha)\Gamma(\alpha+1)}
\int_0^{t_1}dt'\:t'^{\alpha-1}(t_2-t')^\alpha\right\} \nonumber \\
&&+\Theta(t_1-t_2)\left\{\frac{1}{\Gamma(2\alpha+1)}t_2^{2\alpha}+
\frac{1}{\Gamma(\alpha)\Gamma(\alpha+1)}
\int_0^{t_2}dt'\:t'^{\alpha-1}(t_1-t')^\alpha \right\}.
\end{eqnarray}

The integrals can be performed with \textit{Maple} and lead to the
hypergeometric function $F(a,b;c;z)$:
\begin{eqnarray}
\int_0^{t_1} dt'\:t'^{\alpha-1}(t_2-t')^\alpha=\frac{1}
{\alpha}t_1^\alpha t_2^\alpha\:F(\alpha,-\alpha;
\alpha+1;\frac{t_1}{t_2}) \qquad.
\end{eqnarray}

\end{document}